\documentclass[useAMS,usenatbib]{mn2e}
\usepackage{graphicx}
\def\cf{{\it cf.}}
\def\eg{{\it e.g.}}
\def\etal{{\it et al.}}
\def\etc{{\it etc.}}
\def\ie{{\it i.e.}}

\def\reswri{{\it reswri}}
\def\rotcur{{\it rotcur}}
\def\things{{\sc THINGS}}
\def\velfit{{\it velfit}}

\def\vmean{\ensuremath{\bar V_t}}
\def\TdBW{TBWBK}
\long\def\Ignore#1{\relax}

\def\pmb#1{\setbox0=\hbox{$#1$}%
  \kern-0.25em\copy0\kern-\wd0
  \kern.05em\copy0\kern-\wd0
  \kern-0.025em\raise.0433em\box0}

\title[Non-circular Streaming in Disc
Galaxies]{Quantifying Non-circular Streaming Motions in Disc Galaxies}
\author[J. A. Sellwood and R.
Z\'anmar S\'anchez]{J. A. Sellwood$^{1}$\thanks{E-mail:
sellwood@physics.rutgers.edu} and Ricardo
Z\'anmar S\'anchez$^{1,2}$\thanks{E-mail: zanmar@oact.inaf.it} \\
$^{1}$Rutgers University, Department of Physics \& Astronomy, 136
Frelinghuysen Road, Piscataway, NJ 08854-8019, USA \\
$^{2}$INAF -- Osservatorio Astrofisico di Catania, Via Santa Sofia 78,
I-95123 Catania, Italy}
\begin{document}


\pagerange{\pageref{firstpage}--\pageref{lastpage}} \pubyear{2010}

\maketitle

\label{firstpage}

\begin{abstract}
High-quality velocity maps of galaxies frequently exhibit signatures
of non-circular streaming motions.  We here apply the software tool,
\velfit\ recently proposed by Spekkens \& Sellwood, to five
representative galaxies from the \things\ sample.  We describe the
strengths and weaknesses of the tool, and show that it is both more
powerful and yields results that are more easily interpreted than the
commonly used procedure.  We demonstrate that it can estimate the
magnitudes of forced non-circular motions over a broad range of bar
strengths from a strongly barred galaxy, through cases of mild
bar-like distortions to placing bounds on the shapes of halos in
galaxies having extended rotation curves.  We identify mild oval
distortions in the inner parts of two dwarf galaxies, NGC 2976 and NGC
7793, and show that the true strength of the non-axisymmetric gas flow
in the strongly barred galaxy NGC 2903 is revealed more clearly in our
fit to an optical H$\alpha$ map than to the neutral hydrogen data.
The method can also yield a direct estimate of the ellipticity of a
slowly-rotating potential distortion in the flat part of a rotation
curve, and we use our results to place tight bounds on the possible
ellipticity of the outer halos of NGC 3198 and NGC 2403.
\end{abstract}


\section{Introduction}
The centrifugal balance of gas on near circular orbits in a galaxy
yields a direct estimate of the central attraction, which is the first
step towards a model for the mass distribution.  The estimation of
galaxy rotation curves therefore has a long history \citep[see][for a
review]{SR01}.

Well-sampled 2D velocity maps provide sufficient information to
identify the rotation centre and to test whether the line-of-sight
velocity field is, or is not, consistent with a circular flow patttern
in an inclined plane about a common rotation centre.  The widely-used
software utility \rotcur\ \citep{Bege87} divides the velocity map
into a number of elliptical elements that are assumed to be projected
circles around which the gas moves on circular orbits.  It yields an
estimate of the circular speed in each annulus, and has options to fit
for, or hold fixed, the rotation centre, systemic velocity, position
angle and inclination in each annulus.  It is uniquely powerful in its
ability to extract information when the gas layer is warped.

A number of possible systematic errors in the fitted velocity have
been discussed.  Beam smearing \citep{BS01} is obviously reduced by
improved spatial resolution of the observations, \eg, by using optical
data when available.  Various forms of turbulence -- also described as
pressure support or as an asymmetric drift \citep{VR07} -- can be
recognized and corrected for in high quality data \citep{OB08}.

The present paper addresses the systematic error caused by
non-circular flow patterns.  Velocity ``wiggles'' or larger-scale
distortions of the isovelocity contours have long been recognized
\citep[\eg][]{Bosm78} in well-sampled 2D velocity maps.  Both
\cite{HN06} and \cite{VR07} show qualitatively that non-circular
streaming in non-axisymmetric potentials can lead to a mis-estimation
of the central attraction.  Note that \cite{WSW01}, \cite{KSR03},
\cite{PFF4} and \cite{Wein04} had previously published a number of
quantitative models of non-circular flows that took this effect into
account; these papers used the additional information contained in the
non-circular motions of strongly barred or spiral galaxies to separate
the central attractions of the disc and halo.

Not only do non-circular motions arise from bars and other visible
distortions, but they may also be caused by expected asphericities in
dark matter halos \citep[\eg][]{JS02,AFP06,HNS07}.  Because halos are
largely pressure-supported, the principal axes of such distortions can
slew only slowly with radius, and should appear approximately straight
over a moderate radial range.  Even though halo shapes are expected to
be made rounder by disk formation \citep[\eg][]{Dubi94,DM08}, gas in
the outer discs is predicted to flow in an elliptical streaming
pattern in response to forcing by a non-axisymmetric halo.
\cite{HN06} argue that halo-driven non-axisymmetric streaming motions
in the inner disc may mask a central cusp in the halo density.


It is therefore desirable to be able to identify non-circular
streaming motions in high-quality 2D velocity maps and to correct for
their influence when deriving an estimate of the average central
attraction.  The powerful approach developed by \cite{WSW01}
\citep[see also][]{ZSWW} requires multiple datasets and is
time-consuming to implement.  Thus there is a clear need for versatile
tool for routine use.

The procedure proposed by \cite{SFdZ} and \cite{Scho99} estimates the
magnitude of non-circular motions, but does not correct for them.  It
is embodied in the tool \reswri, which is an extension of \rotcur,
that has been used quite extensively.  More recently
\citet[][hereafter SS07]{SS07} proposed an alternative tool, \velfit,
that does provide an improved estimate of the mean orbital speed.
Here we demonstrate the superior performance of \velfit\ by direct
comparison of the results from several galaxies.  We have chosen to
make this comparison using a few galaxies from the
\things\ \citep{WBdB} sample, for which non-circular motions have
recently been estimated using \reswri\ by \citet[][hereafter
  \TdBW]{TdBW}.  In addition, we show that \velfit\ can also be
applied to optical data from a strongly barred galaxy in the
BH$\alpha$Bar \citep{BHaB} galaxy sample.

\section{Estimating Non-circular Motions}
\label{better}
In this section, we describe and compare the two separate software
tools that are available for estimating the magnitudes of non-circular
flows.

The procedure proposed by \cite{SFdZ}, embodied in the tool \reswri,
is an extension of \rotcur\ to include an harmonic analysis of the
line-of-sight velocities around each ring.  These authors use epicycle
theory to interpret the fitted non-axisymmetric coefficients, and
relate small values to the magnitude of the potential distortion.

It should be noted that non-circular motions are readily confused with
the kinematic signature of a warp, since both cause variations in the
ellipticity and position angle of the flow pattern.\footnote{In fact,
  the kinematic signatures of an oval potential and a warp are
  degenerate only when the principal axis of the potential coincides
  with either the major or minor axis of the projection
  \citep{PB84,FvGZ}.}  Thus, if the projection geometry is allowed to
vary from ring to ring, then a large part of the actual non-circular
motion may be masked by radial variations in the position angle (PA)
and inclination ($i$).  Thus most users \citep[\eg][]{Fathi05,vETK}
advocate constraining $i$ \& PA to have the same values at all radii.
On the other hand, \TdBW\ justified allowing individual tilts for
rings in all the \things\ galaxies from the fact that the magnitudes
of non-circular streaming motions in two galaxies (NGC~3198 \&
DDO~154) were little changed when $i$ \& PA were allowed to vary
compared with when they were constrained to be constant with radius.
Here we show that allowing rings to tilt independently led them to
miss non-axisymmetric distortions in at least two galaxies: NGC~2976
\& NGC~7793.

The tool \reswri\ has a number of disadvantages.
\begin{enumerate}
\item A $\cos(m\theta)$ component in the projected velocities can
  arise from both $\cos[(m\pm1)\theta]$ distortions to the potential
  \cite[][see also SS07 and references therein]{SFdZ}, complicating
  the interpretation of the results.
\item It assumes the perturbed velocities are small, and is therefore
  inadequate to characterize strongly non-axisymmetric flows.
\item It makes no correction for the presence of non-circular motions.
  As a result, the estimated circular speed, $V_c(R)$, is biased
  toward the value of the azimuthal speed on the major axis of the
  projection, as explained below.
\item It treats each ring independently, implying that a mild
  distortion that is coherent over a significant radial range is more
  easily masked by noise.
\item An estimate of the strength of the mild potential distortion
  responsible for the detected non-axisymmetric flow requires an
  undesirable difference between two of the fitted coefficients, a
  consequence of point (i) above, and also includes the sine of an
  angle whose value cannot be determined from this approach.
\end{enumerate}

\begin{figure}
\centerline{\includegraphics[width=\hsize]{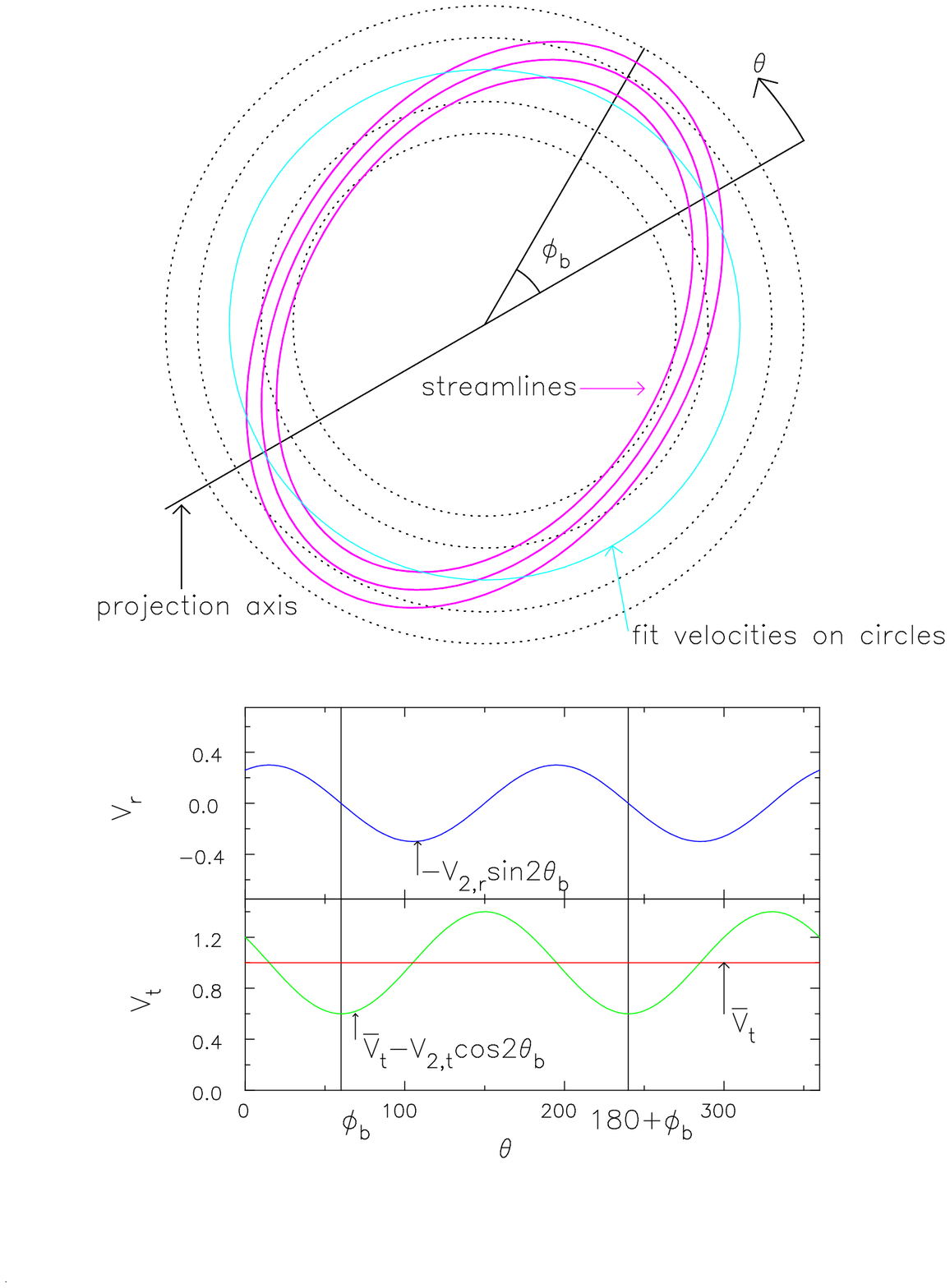}}
\caption{A schematic illustration of the approach taken by
  \citet{SS07}.  The magenta lines in the upper panel indicate
  non-circular streamlines in the disc plane, which is shown face-on.
  The galaxy is observed in projection, with the intersection of the
  sky-plane and the disc plane making an angle $\phi_b$ to the major
  axis of the elliptical streamlines.  Our model of the flow consists
  of a set of values for \vmean, $V_{2,t}$ and $V_{2,r}$ around rings
  (dotted) at fixed radii.  We project the model and fit to the
  observed data, using linear interpolation to predict data values
  between rings. \hfil\break The lower panels illustrate the bisymmetric
  variations of the radial and azimuthal model velocities (arbitrary
  scale) in the disc plane around the cyan circle in the upper panel.
  The radial velocity (blue) varies with angle $\theta_b = \theta -
  \phi_b$ to the major axis of the elliptical streaming pattern as
  $-V_{2,r}\sin2\theta_b$.  The azimuthal velocity (green) varies as
  $-V_{2,t}\cos2\theta_b$ about the mean (\vmean\ shown in red).  This
  physically-motivated phase difference between the two
  non-axisymmetric components is embodied in the fit.}
\label{schematic}
\end{figure}

SS07 therefore proposed an alternative tool, \velfit\ illustrated in
Fig.~\ref{schematic}.\footnote{The \velfit\ software is publicly
  available from {\tt
    http://www.physics.rutgers.edu/$\sim$spekkens/velfit/}.} It
differs at root by postulating a specific model of the flow that
includes a possible non-circular streaming pattern about a fixed
direction in the disc plane, \ie, a distortion that has no spirality.

In particular, SS07 assume a flat disc plane and a distortion having a
fixed orientation at all radii, although its amplitude may vary with
$R$.  The tool \velfit\ then fits the projected model velocity at a
general point (eq.\ 5 of SS07):
\begin{eqnarray}
\nonumber
V_{\rm model} & = & V_{\rm sys} + \sin i \;\left[\vmean\cos\theta
\right. \\
\nonumber
&& \qquad - \; V_{m,t}\cos m(\theta-\phi_b) \cos\theta\\
&& \qquad \left. - V_{m,r}\sin m(\theta -\phi_b) \sin\theta \,\right],
\label{pveleq}
\end{eqnarray}
to derive $\vmean(R)$, $V_{m,t}(R)$, $V_{m,r}(R)$, the angle $\phi_b$,
the systemic velocity $V_{\rm sys}$, and $i$ \& PA.  (The various
quantities are defined in the caption to Fig.~\ref{schematic}, which
illustrates an elliptical streaming pattern for which $m=2$.)

The code fits a model by minimizing the usual function
\begin{equation}
\chi^2 = \sum_{n=1}^N \left( {V_{\rm obs}(x,y) - \sum_{k=1}^K w_{k,n}
  V_k \over \sigma_n} \right)^2.
\label{chisq}
\end{equation}
Here $V_{\rm obs}(x,y)$ and $\sigma_n$ are respectively the value of
the observed velocity and its uncertainty for the $n$-th pixel at the
position $(x,y)$ on the sky.  The index $k$ ranges over $K$, which is
the total all three sets of velocities, $V_k$, that define the model
velocity (eq.~\ref{pveleq}) around each of the ellipses.  The weights
$w_{k,n}$, which include the trigonometric factors, also define the
interpolation scheme that yields a model prediction at the projected
position $(x,y)$.  The code allows the radial extent of the distortion
to be restricted, if desired, while a simple axisymmetric flow is
fitted over the remainder of the data.  SS07 describe the algorithm in
detail.

We here use the term ``bar model'' to denote any straight bisymmetric
distortion no matter what its origin or amplitude.
Fig.~\ref{schematic} should not, however, be interpreted to imply that
the code can fit only bars, or oval features, although that is its
most useful application.  The same code can be used to fit models
having higher or lower rotational symmetry, $m$, or axisymmetric
($m=0$) radial flows.

This tool avoids the above-listed disadvantages of the method devised
by \cite{SFdZ}, as follows.
\begin{enumerate}
\item Fitting a specific distorted flow model to the projected data
  avoids the complications caused by the coupling of different angular
  periodicities.
\item It can fit for arbitrarily large distortions because it does not
  require $V_{m,t}(R)$ \& $V_{m,r}(R)$ to be related by the epicycle
  approximation.
\item It yields \vmean, which is an improved estimate of the average
  orbital speed at each radius, as discussed below.
\item It uses all the data in a single fit, making it easier to
  identify coherent mild distortions in noisy data and to go some way
  towards ``averaging over'' small-scale spiral streaming.
\item The magnitude of a mild potential distortion is much more
  directly related to the fitted velocity coefficients, as we show in
  section~\ref{epic}.
\end{enumerate}
An elliptical flow pattern is clearly an idealization.  It is a
first-order improvement that captures the essential large-scale
features of a bar-like flow, and will therefore be a better fit than a
flat, axisymmetric model.  It does not attempt to include other
features, such as shocks in the bar, spiral arms, turbulence, warps,
\etc, that are also generally present in the data.

The assumption of a flat disc plane is equivalent to requiring
constant $i$ \& PA in \reswri, which is the usual practice, except
that \velfit\ has the further advantage that it determines the optimal
values as part of the fit.  The inner parts of spiral galaxy discs are
believed to be flat: warps are generally observed to start near the
edge of the optical disc \citep[\eg][]{Brig90} and theoretically we
expect the massive inner disk to be coherent enough to resist bending
\citep[\eg][]{SS06}.

With the projection angles held at constant values for all radii, it
may seem from Fig.~\ref{schematic} that fitting a simple circular flow
model, \eg\ with \rotcur, would yield the same \vmean\ as a
bisymmetric model from \velfit.  However, the estimated circular
speeds for an axisymmetric model are biased towards values on the
major axis where the line-of-sight component of the orbital motion is
greatest.  Since gas moving on elliptical orbits has its smallest
orbital speed at apocentre,\footnote{Even in strong bars, where
  angular momentum is not approximately conserved, gas at apocentre of
  the stream lines must be moving slower than the average since it
  plunges inwards after that point, irrespective of the angular
  rotational speed of the potential that drives the non-circular
  flow.}  the fitted \vmean\ is biased low when the streaming pattern
is closely aligned with the major axis.  Conversely, we should expect
an axisymmetric fit to be biased high by the higher-than-average speed
of the gas near pericentre when the bar is oriented close to the
projected minor axis, and to yield a fair estimate of \vmean\ when the
bar is at intermediate angles.  Recall that \vmean\ is defined to be
the average tangential velocity around a circle that may cross
multiple streamlines, as shown in the strongly non-axismmetric flow
sketched in the upper panel of Fig.~\ref{schematic}.  Thus even though
\vmean\ affords a fairer estimate of the central attraction than that
from an axisymmetric fit, it is not the circular speed in the
equivalent axisymmetric potential, unless the distortion is very mild.

Because \reswri\ fits each ring independently, it is not easy to apply
a smoothing constraint to the fit.  Spiral arms, turbulence, \etc,
produce localized distortions to the flow that can lead to ``wiggles''
in the fitted velocities, as well as small-scale variations in $V_{\rm
  sys}$, $i$, \& PA.  The user of \reswri\ can, and probably should,
hold these global parameters fixed, but the tool does not have the
option to smooth the fitted velocities.  While excessive smoothing is
clearly dangerous, \eg\ it could reduce the slope of the inner
rotation curve, a small degree of smoothing applied to constrain the
fit can be beneficial. In Appendix A we describe how an optional
penalty can be applied within \velfit\ to smooth the radial variations
of the fitted functions $\vmean(R)$, $V_{m,t}(R)$, \& $V_{m,r}(R)$;
the magnitude of the penalty can be set independently for the
axisymmetric and non-axisymmetric terms.

The tool \velfit\ assumes a straight position angle for the
non-axisymmetric distortion, which has both advantages and
disadvantages.  Since it cannot follow the radial winding of a spiral
pattern, it ``averages over'' the spirals (advantage (iv) mentioned
above) and fits only straight distortions with $m$-fold rotational
symmetry.  Its primary purpose therefore is to identify bar-like or
oval distortions in the disk or halo.  However, a fixed position angle
precludes mapping the distorted flow of a spiral, which could be an
interesting capability.  It may be possible to adapt \velfit\ to
include extra parameters to fit a spiral of a certain shape, which we
leave for possible future development.

A weakness of \velfit\ is that it tends to return absurd velocities
when the bar angle, $\phi_b$, is near zero or $90\degr$ because a
degeneracy arises between the velocity components at these special
orientations.  To see this, consider equation (\ref{pveleq}) for the
predicted projected velocity when $\phi_b=0$: the products
$\cos2\theta\cos\theta$ and $\sin2\theta\cos\theta$, can be separated
into a part that varies as $\cos\theta$ and another that varies as
$\cos3\theta$ or $\sin3\theta$.  Therefore \vmean\ is partly
degenerate with both $V_{2,t}$ and $V_{2,r}$, and a similar partial
degeneracy arises when $\phi_b = 90\degr$.  In principle, the
$3\theta$ variation of the model breaks the degeneracy, but these more
rapid angular variations are more susceptible to noise, and the ``best
fit'' values of the three velocity components can be absurd.  We show
an example in section \ref{strong}, where smoothing (Appendix A)
proves valuable in controlling this numerical artefact.

The \velfit\ code can be used to fit simultaneously for more than one
type of distortion, but we have not obtained anything useful from
attempting this.  However, \cite{vdVF} suspect both spiral
perturbations and axisymmetric radial inflow in the central parts of
NGC~1097, and fit for both using a generalization of the method
proposed by \cite{SFdZ}.

It should be noted that \velfit\ is purely a fitting procedure.  It
does not, in general, yield a direct estimate of the potential
responsible for the fitted velocity distortions, except when
departures from axial symmetry are mild, as we show in the next
section.

\section{Mildly Distorted Potentials}
\label{epic}
Where the fitted non-circular speeds are a small fraction of the
circular speed, we can use epicycle theory to relate the fitted
non-circular velocities to the strength of the non-axisymmetric
perturbation.  We can use the formulae for the orbits of test
particles on near circular orbits in weakly barred potentials from
\citet[][hereafter BT08]{BT08}.\footnote{Gas streamlines trace test
  particle orbits when pressure and magnetic forces can be neglected.}
Their equation (3.147a, p190) gives the forced radial displacement as
a function of time, which can easily be differentiated to find the
forced radial speed.  The radial velocity at radius $R_0$ varies
sinusoidally with amplitude
\begin{equation}
V_{m,r} = \left[{d\Phi_b \over dR} + {2\Omega\Phi_b \over R(\Omega -
    \Omega_b)} \right]_{R_0}{m(\Omega_0 - \Omega_b) \over \kappa_0^2 -
  m^2(\Omega_0 - \Omega_b)^2},
\label{vrepi}
\end{equation}
where $\Phi_b$ is the weak non-axisymmetric part of the potential that
rotates at angular rate $\Omega_b$, and $\Omega(R)$ \& $\kappa(R)$ are
the usual frequencies of rotation and epicycle motion for mildly
eccentric orbits.  The time derivative of the equation for the
tangential displacement \citep[\eg][eq.~10b]{SW93}, converted to the
same notation and abbreviating $\omega = m(\Omega-\Omega_b)$, gives
\begin{equation}
V_{m,t} = \left[{2\Omega \over \omega}{d\Phi_b \over dR} + {4\Omega^2
    -\kappa^2 + \omega^2 \over \omega^2}{m\Phi_b\over R} \right]_{R_0}
{\omega_0 \over \kappa_0^2 - \omega_0^2}.
\label{vtepi}
\end{equation}

With aspherical halos in mind, we simplify these general expressions
for the case of a non-rotating potential distortion in a region where
the rotation curve is approximately flat.  (Other assumptions are
possible, if desired.)  Since we expect the halo to be rotating
slowly, we assume $\Omega_b \ll \Omega_0$, we set $\kappa_0^2 =
2\Omega_0^2$ for a flat rotation curve, and we assume the potential
perturbation varies slowly with radius so that $d\Phi_b/dR \ll
\Phi_b/R$.  With these assumptions, choosing $m=2$ for a bisymmetric
distortion and setting $R_0\Omega_0 = V_c$, equations (\ref{vrepi})
and (\ref{vtepi}) reduce to
\begin{equation}
V_{2,r} \simeq -{2\Phi_b \over V_c}, \qquad \& \qquad
V_{2,t} \simeq -{3\Phi_b \over V_c}.
\label{amp}
\end{equation}
Thus if the perturbed velocities are caused by a weak, non-rotating,
oval distortion to a quasi-logarithmic potential, we should expect the
perturbed velocity coefficients to be in the ratio $V_{2,t} \simeq
1.5V_{2,r}$ and, in particular, they should have the same sign.  Note
that we do not expect the perturbed coefficients to have this ratio
when the oval distortion is strong and/or rapidly rotating, such as
for bars, or in the inner parts where the rotation curve is rising.

In order to relate $\Phi_b$ to the potential shape, we assume a
non-axisymmetric potential of the form (\cf\ eq.\ 2.71a of BT08)
\begin{equation}
\Phi(R,\theta) = {V_0^2 \over 2} \ln\left[ 1 + {R^2 \over R_c^2}
  \left( 1 + {1-q_\Phi^2 \over q_\Phi^2} \sin^2\theta \right)\right],
\end{equation}
where $V_0$ sets the velocity scale, $R_c$ is the core radius, and
$q_\Phi$ is the axis ratio of the potential.  In the limit of $q_\Phi
\la 1$, this is the potential of a midly non-axisymmetric galaxy
having flat outer rotation curve, of the kind we assumed above.
Expansion of this potential for small $(1-q_\Phi^2)/q_\Phi^2$, and
comparison with the definition of $\Phi_b$ in equations (3.136) and
(3.143) of BT08, we find $\Phi_b = -V_c^2(1-q_\Phi^2) /(4q_\Phi^2)$.
Combining this result with eq.~(\ref{amp}) and equating \vmean\ to
$V_c$, we finally obtain
\begin{equation}
q_\Phi = \left({\vmean \over \vmean + 2V_{2,r}}\right)^{1/2}.
\label{potshape}
\end{equation}
We stress that this formula assumes a mildly distorted, slowly
rotating potential and a flat rotation curve, and it will not yield a
reliable estimate in other circumstances.  As is well known, the
density that gives rise to this potential is about three times more
elongated than the potential, so that $q_\rho \simeq 1 - 3(1-q_\Phi)$
(BT08, p.~77).

\section{Results}
Here we apply the tool \velfit\ to several galaxies in the
\things\ sample \citep{WBdB}.  The HI Nearby Galaxy Survey (\things)
used the Very Large Array operated by the National Radio Astronomy
Observatory\footnote{NRAO is a facility of the National Science
  Foundation operated under cooperative agreement by Associated
  Universities Inc.} to make spectral observations of the 21cm line
emission of neutral hydrogen in a sample of 34 galaxies.  The data are
in the public domain.  We do not reanalyse the entire \things\ sample
here, but choose a few representative galaxies to illustrate the
advantages of \velfit\ over \reswri.

We selected data with natural weighting, which have higher
signal-to-noise (S/N) and lower spatial resolution than when robust
weighting is used, and downloaded maps of the intensity-weighted mean
velocity.  In order to apply a S/N cut-off, we also downloaded the
data cubes.  We determined the noise level, $\sigma$, from parts of
channel maps with no signal, and discarded velocity measurements from
the maps for which the peak intensity in any channel $< 5\sigma$.

The usual $\chi^2$ function, defined in eq.~(\ref{chisq}), requires an
estimate for the error $\sigma_n$ in the data at each point in the
map.  Since the neutral hydrogen clouds in a galaxy have typical
random velocities of 6-12~km~s$^{-1}$ \citep{Kamp}, we do not adopt
the formal (generally much smaller) uncertainties in the
intensity-weighted mean velocities, but instead assume a constant
uncertainty of $\sigma_n = 10\;$km~s$^{-1}$.  We have checked that our
results are insensitive to values in the range $\sigma_n =
10\pm4\;$km~s$^{-1}$.

We choose the spacing between the rings in our model to be the beam
width in order that each ring is independent.  The beam size differs
for each galaxy, ranging from $7.41\arcsec$ for NGC~2976 to
$15.6\arcsec$ for NGC~7793.  Since smoothing carries with it the
danger that it may ``smooth away'' real features in the rotation
curve, we do not employ it as a matter of routine.  However, smoothing
is essential to stabilize the fits to the data for NGC~2903, where the
bar is close to the major axis (see \S\ref{better}).

We adopt the distances given in Table 1 of \cite{WBdB} solely for the
purpose of marking radii in kpc as well as in arcsec in the rotation
curve plots.

\subsection{Errors}
Velocity residuals at each pixel have correlations due to features in
the data, caused by spiral arms for example, that are not part of the
fitted model.  In addition, a small fraction of pixels have large
residuals.  These properties of the data, together with our use of
arbitrary uncertainties, imply that the values on the $\chi^2$ surface
cannot be used to obtain error estimates on the parameters of the
model.

We therefore adopt a bootstrap technique to estimate the true
uncertainties in all fitted quantities, including the velocities.
This standard non-parametric method \citep[\eg][]{Cher} yields
statistical uncertainties in fitted quantities without making
assumptions about the underlying distribution of the residuals between
the model and the data values.  The spread of estimated quantities
from repeated fits to resampled (or pseudo-)data yields an estimate of
the true uncertainty.  We give further details of our procedure
in Appendix B.

The error bars on our velocity estimates, and the uncertainties in our
tabulated values, are the rms variations of each quantity from 1000
bootstrap iterations.  These statistical uncertainties reflect the
peculiarities in the distribution of the errors in the data values, as
well as some of the systematic inadequacies of the fitted model, which
ignores features such as shocks or spiral streaming.  They do not
include possible systematic errors from other sources such as beam
smearing or pressure support, neither can they reflect possible
systematic differences between different data sets.  As we report
here, different kinematic data obtained from a different instrument,
and perhaps arising from a different component of the ISM, can yield
different estimates of the same physical quantities.  A disagreement
by more than the properly estimated errors is an indication either of
systematic errors in one or both data sets or the inadequacy of the
model.

\subsection{NGC 2976: A Case Study}
NGC 2976 is a nearby dwarf galaxy in the \things\ sample which was
also studied in great detail by \cite{SBLB}.  Velocity maps have
therefore been made using the 21 cm line of HI, the optical H$\alpha$
emission lines, and the $^{12}$CO($J=1 \rightarrow 0$) line.  Clear
distortions are visible in the velocity maps, indicating departures
from a simple, coplanar axisymmetric flow.

\begin{figure}
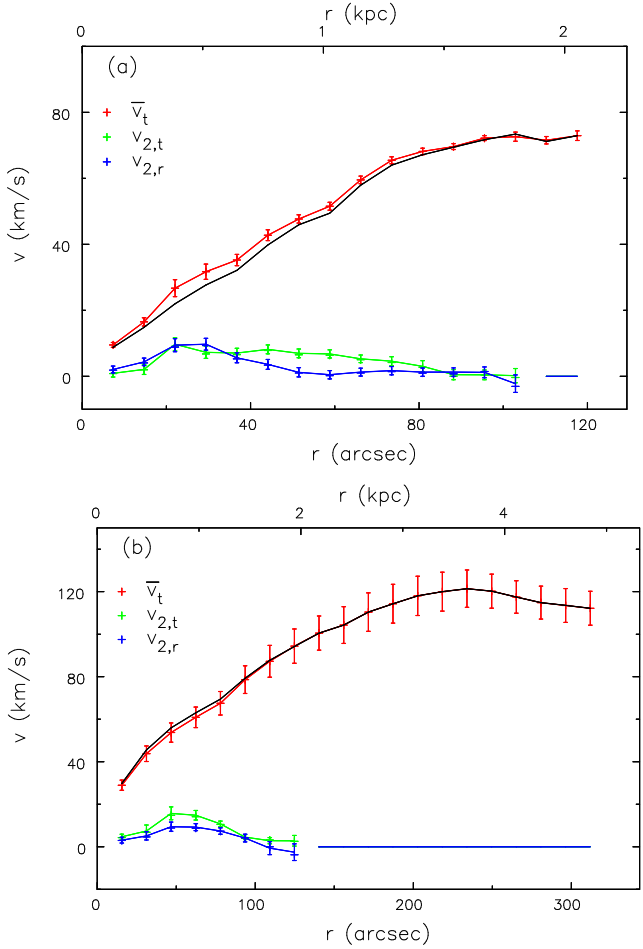

\includegraphics[width=.98\hsize,angle=0]{n2976_rc.ps}
\par\vspace{.2cm}
\includegraphics[width=\hsize,angle=0]{n7793_rc.ps}
\caption{Our best fit rotation curve and streaming velocities derived
  using \velfit\ on the \things\ data for (a) NGC~2976 and (b)
  NGC~7793.  The black line shows the result of an axisymmetric fit to
  all radii and the error bars represent $\pm\sigma$ uncertainties.}
\label{n2976}
\end{figure}

\cite{SBLB} fitted their data with a model that combined the usual
circular flow pattern with an axisymmetric radial flow.  SS07 found
that an oval or bar-like distortion could yield an equally good fit to
the same data.  \TdBW\ find that a tilted ring model fits the HI
data, which they conclude is consistent with tiny deviations from a
round potential, albeit with a $\la 30\degr$ variation in the
inclination and a similar change of PA in the inner part of the
galaxy.

Fig.~\ref{n2976}(a) shows our fit to the same HI observations used by
\TdBW\ when we assume a flat disc plane and fit for bisymmetric flows
to $R=105\arcsec$ only.  The residual velocities after subtracting our
best fit model are typically less than 5 km/s, and fewer than 1\% of
the pixels have residuals larger than 20 km/s.  The error bars show
$\pm\sigma$ uncertainties from the bootstrap analysis. \cite{dBWB}
report a mild warp for $R \ga 125\arcsec$, but we have excluded data
at these large radii because they have low S/N.  While our estimate of
PA (Table~\ref{n2976_pars}) is in good agreement with that derived
from the same HI data by \cite{dBWB}, our value of the inclination is
somewhat lower than the $i=65\degr$ they adopted, although their value
is estimated from the restricted annular range $80\arcsec \leq R \leq
110\arcsec$, whereas ours is from a global fit that includes
bi-symmetric streaming over most of the disc.

\begin{table}
\caption{Best-fit parameters for NGC 2976 and NGC 7793}
\label{n2976_pars}
\begin{tabular}{@{}lrr}
                   & NGC 2976       & NGC 7793 \\
\hline
Adopted distance (Mpc)  &   3.6     &   3.9 \\
Systemic vel. (km s$^{-1}$)      &  $1.6 \pm 0.1$ &  $226.9 \pm 0.1$ \\
Disc inclination $i$    &  $ 61\degr \pm 1\degr$ &  $ 44\degr \pm 5\degr$ \\
Disc PA, $\phi_d^\prime$ &  $323\degr \pm 1\degr$ &  $292\fdg0 \pm 0\fdg4$ \\
Bar axis $\phi_b$       &  $-29\degr \pm 6\degr$ &  $ 51\degr \pm 8\degr$ \\
Projected bar axis $\phi_b^\prime$ &  $308\degr \pm 3\degr$ &  $334\degr \pm 7\degr$ \\
\end{tabular}
\end{table}

We find (Fig.~\ref{n2976}a) clear non-circular streaming motions in
the inner galaxy in these HI data that are similar to, but of somewhat
smaller magnitude than, those found by SS07 from the combined
H$\alpha$ and CO data.  Even though the perturbed velocities are $\la
10\;$km~s$^{-1}$, they are still a large enough fraction of the orbit
speed that \velfit\ is required; notice that the black line, which
shows $V_c$ estimated in an axisymmetric fit such as would be found
from \reswri, is consistently lower than $\vmean$.  We also find a
smaller inclination angle than $i\sim64\degr$ fitted by SS07.  The
origin of these differences, which are barely consistent within the
quoted errors, is clearly due to fitting different datasets, with HI
data arising from a different physical component.

We estimate the projected orientation of the bar or oval to be
$\phi_b^\prime \simeq 308\degr\pm3\degr$, which is barely consistent
with the $\sim 315\degr\pm4\degr$ estimated (from different data) by
SS07.  The difference in angles implies that the bar is slightly
farther from the major axis, which in turn reduces the difference
(Fig.~\ref{n2976}a) between \vmean\ and the axisymmetric fit, for the
reason given in section \ref{better}.  Note, however, that we obtain a
projected bar angle in closer agreement with that estimated by SS07 if
we fix the galaxy inclination at their estimated value, indicating
that the non-circular streaming motions in the neutral hydrogen arise
from forcing by the same non-axisymmetric potential.

The probable reason that the bar streaming velocities we find here are
weaker than those of SS07 is the significantly less pronounced
`S'-shaped distortions of the velocity contours in the HI data (see
Fig.~28 of \TdBW) than those shown in Fig.~2(a) of SS07.  Since the
distortions are well-resolved, it is unlikely that the differences
result from the minor difference in spatial resolution of the two
datasets: the beam-size of the HI data, $7.41\arcsec$, is $\sim 50$\%
greater than the $5\arcsec$ smoothing length adopted by \cite{SBLB}
for the combined CO and H$\alpha$ data used by SS07.  (Note that
\cite{SBLB} show that velocity estimates from both CO and H$\alpha$
data, which are almost entirely from HII regions, are generally in
good agreement with each other.)  We can only speculate as to why the
neutral hydrogen has milder non-circular motions than either the
molecular or ionized gas; perhaps the HI layer has a greater physical
thickness than the molecular/ionized layer and therefore greater
``pressure'' (in reality larger turbulent speeds), which causes a
weaker response to a non-axisymmetric potential.  Note that if the
source of the non-axisymmetric potential is a bar in the disc of this
galaxy, a thick gas layer will feel a weaker potential only if the bar
is much thinner vertically than is the HI layer.

Since all three models, with radial flows \citep{SBLB}, a twisted disc
(\TdBW), or oval streaming (SS07 and Fig.~\ref{n2976}a), are adequate
fits to the data, there is no statistical reason to prefer one over
another.  However, the oval streaming model both avoids the
``continuity problem'' inherent in radial flow models, and also avoids
a strong twist in the plane of the inner disc; \TdBW\ suggest the disc
plane at $R\sim 20\arcsec\;$($\simeq 300\;$pc) is inclined to the
plane of the main part of disc at $R \sim 1.5\;$kpc by $\sim 30\degr$,
while the light distribution does not give any indication of such an
unusual feature.

\subsection{Effects of bar orientation}
The value of \vmean\ in Fig.~\ref{n2976}(a), while smaller than
obtained from different data by SS07, is higher than results from a
purely axisymmetric flow fit, shown by the black line.  The
bisymmetric fitted \vmean\ is larger in this case because the ``bar''
is oriented such that its principal axis is not far from to the major
axis of projection for the galaxy, as discussed in
section~\ref{better}.  We therefore also present a case where the
bisymmetric fit has little effect on \vmean.

Fig.~\ref{n2976}(b) shows a fit to the \things\ data for NGC~7793.  As
always, we attempt to fit only the flat part of the disc; in this
galaxy, \cite{dBWB} find a mildly varying disc inclination over the
entire radial range, but the PA clearly rises steadily for $R \ga
320\arcsec$, which we tentatively interpret as the start of a warp.
We fit a bisymmetric flow over the inner disk $R<125\arcsec$ only and
an axisymmetric model to $R = 320\arcsec$.  The best fit parameters
and uncertainties are given in Table~\ref{n2976_pars}.  Our best-fit
$i$ \& PA are in good agreement with the values estimated by
\cite{dBWB} from their tilted ring analysis.

The uncertainty in our estimate of the inclination is, however, rather
larger than in other cases, possibly because of spiral streaming in
the outer disc, which is not included in our model, or perhaps because
the entire disk is warped, as suggested by \TdBW.  Their explanation
seems the more likely because the bootstrap values for this parameter
have a distinctly bimodal distribution symmetrically distributed about
the best-fit value; conservatively, we estimate the uncertainty from
the rms spread of the bootstrap iterations.  The uncertainty in the
inclination is reflected in the uncertainties in the velocities, which
therefore seem large relative to the smoothly varying means.

We find clear evidence (Fig.~\ref{n2976}b) for non-circular streaming
in the inner parts.  Since we find $\phi_b \simeq 49\degr$, the
estimated ``rotation curve'' from the bisymmetric fit is in close
agreement with that from the simple circular flow model, as expected
from the discussion in section~\ref{better}.  \TdBW\ describe this
galaxy as their best candidate for a non-axisymmetric potential in the
{\it outer parts\/}, but they acknowledge that the problem of the
apparent changing inclination complicates this interpretation.
However, the more rapid radial changes in $i$ \& PA in the inner parts
of their fits again appear to have masked the non-circular streaming
we detect.

\begin{figure}
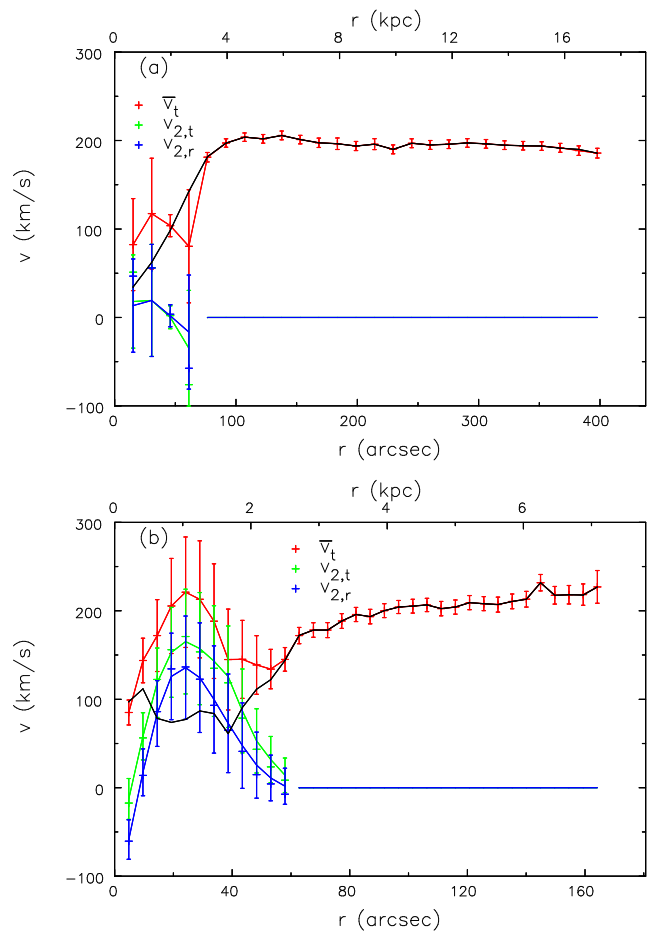

\includegraphics[width=\hsize,angle=0]{n2903a_rc.ps}
\par\vspace{.2cm}
\includegraphics[width=\hsize,angle=0]{n2903b_rc.ps}
\caption{Our best fit rotation curve and streaming velocities derived
  for NGC~2903 using \velfit\ on (a) the \things\ data and (b) the
  BH$\alpha$Bar data -- note the different scales.}
\label{n2903}
\end{figure}

\subsection{Strong bars}
\label{strong}
\TdBW\ find non-circular motions that are consistent with a round
potential in all the \things\ galaxies (their Table 3), despite the
fact their sample contains several galaxies that are quite strongly
barred.  In these cases, however, the 21cm line emission from the
barred region is generally too weak to yield reliable velocity
estimates; velocities can be measured in limited patches in some
cases, \eg\ NGC~925, but not anywhere in others, \eg\ NGC~3627.
However, the data from NGC~2903 are good enough over almost the entire
barred region ($R \la 60\arcsec$) to yield velocities above our S/N
threshold.  In this galaxy, \TdBW\ estimate the median value of $m=2$
streaming motions in the bar region is only $\sim 14\;$km~s$^{-1}$,
and conclude that the mean elongation of the overall potential is
nevertheless consistent with being round.

The application of \velfit\ to this galaxy presents a substantial
challenge because the bar is so closely aligned with the major axis of
projection.  As discussed in section \ref{better}, the velocities
\vmean, $V_{2,t}$ \& $V_{2,r}$ become harder to distinguish as
$|\phi_b| \rightarrow 0$, and \velfit\ can return unphysically large
values for all ($\gg 1000\;$km~s$^{-1}$ in magnitude), even though
they still fit the data well when combined as eq.~(\ref{pveleq}).  We
overcome this problem to a large extent by applying a very small
smoothing penalty, as described in Appendix A.  Fig.\ref{n2903}(a)
shows the results from \velfit\ applied to the \things\ data for
$R<400\arcsec$; the parameters of this fit are given in
Table~\ref{n2903_pars} for $\lambda = 1.7 \times 10^{-4}$, but results
are insensitive to variations of a factor of a few about this
value.\footnote{This apparently bizarre choice is $\lambda=0.002/\cal
  A$ for $V_{\rm typ}=200\;$km s$^{-1}$; see Appendix A.}

\begin{table}
\caption{Best-fit parameters for NGC 2903}
\label{n2903_pars}
\begin{tabular}{@{}lrr}
                   & \things\ data       & BH$\alpha$Bar data \\
\hline
Adopted distance (Mpc)  &   8.9 \\
Systemic vel. (km s$^{-1}$) &  $549.9 \pm 0.3$ &  $554.1 \pm 0.5$ \\
Disc inclination $i$    &  $ 64\degr \pm 1\degr$ &  $ 66\degr \pm 3\degr$ \\
Disc PA, $\phi_d^\prime$ &  $201\fdg5 \pm 0\fdg5$ &  $204\degr \pm 1\degr$ \\
Bar axis $\phi_b$       &  $ 6\degr \pm 14\degr$ &  $-12\degr \pm  8\degr$ \\
Projected bar axis $\phi_b^\prime$ &  $204\degr \pm 6\degr$ &  $199\degr \pm 4\degr$ \\
Smoothing penalty, $\lambda$ & $1.7\times 10^{-4}$ & $6.6 \times 10^{-5}$ \\
\end{tabular}
\end{table}
 
The smoothing penalty successfully eliminates absurd velocities in the
bar region obtained from the bootstrap analysis, but at the cost of
introducing a strong bias against finding $|\phi_b| \la 2\degr$.  The
reason appears to be a ridge in the $\chi^2$ surface as $\phi_b
\rightarrow 0$ caused by the smoothing penalty, which disfavours the
wildly varying velocities that would achieve the smallest residuals.
We therefore prefer a very small smoothing penalty since larger values
widen the range of disfavoured bar angles even though they further
reduce the scatter of fitted velocities from the bootstrap iterations.
On the other hand, increasing the smoothing penalty 50-fold, leads to
an almost linear rise in \vmean, even for the bisymmetric fit, which
illustrates the perils of oversmoothing.

\begin{figure*}
\includegraphics[width=.47\hsize,angle=270]{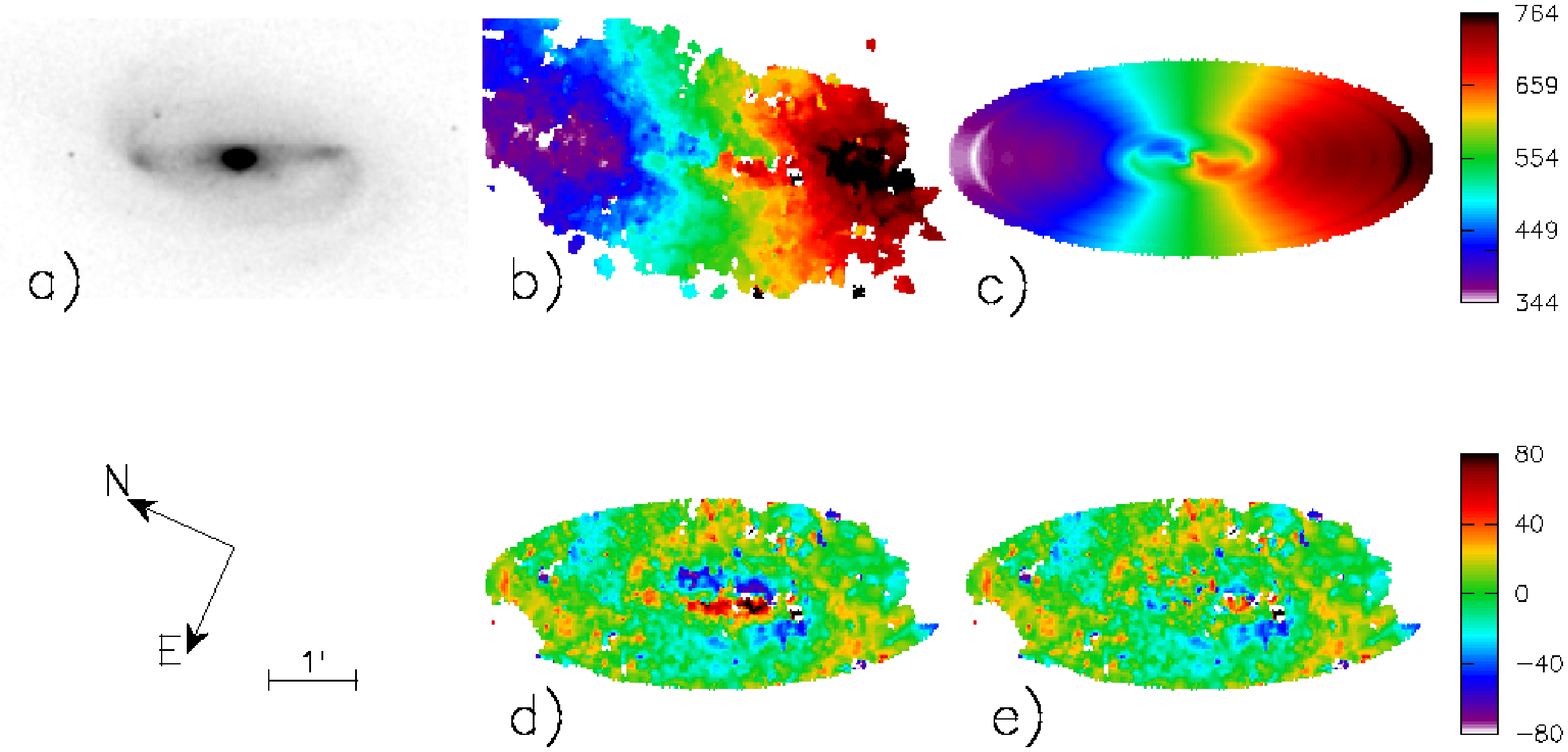}
\caption{(a) An H-band image of NGC 2903 from 2MASS \citep{2MASS}. (b)
  The velocity map made using the H$\alpha$ line \citep{BHaB}. (c) Our
  best fit model with a bar flow. (d) Residuals after subtracting the
  best fit axisymmetric model -- note the large values in the bar
  region. (e) Residuals after subtracting our best fit bi-symmetric
  flow model.  The velocity scales to the right are in km s$^{-1}$;
  all images have the same orientation and spatial scale, indicated by
  the $1\arcmin$ scale bar.}
\label{n2903fig}
\end{figure*}

Our best-fit estimate of the bar angle is $\phi_b \sim 6\degr$. It
should be noted that neither this angle, nor the best fit values of
\vmean\ change significantly when we eliminate the smoothing penalty
altogether.

Our fitted values of $i$ \& PA are in good agreement with those
estimated by \cite{dBWB}, and their rotation velocities are in good
agreement with our axisymmetric fit, the black line in
Fig.~\ref{n2903}(a), that rises in a quasi-linear fashion from the
origin.  Naturally, they find wildly varying values of both $i$ and PA
in their innermost few rings, whereas \velfit\ requires a flat plane
and consequently our bisymmetric fit (red line) finds different
velocities in this region.  We find somewhat larger non-circular
speeds in the bar region than those estimated by \TdBW, although the
uncertainties in our estimates also are large enough that a round
potential could not be excluded.  Thus, the surprisingly weak bar
streaming motions are not merely an artefact of \reswri.

Since the beam width of the HI data we fit is $\sim 15\arcsec$ (equal
to our ring spacing in \velfit) and the bar semi-axis is about four
beam widths, it is likely that the HI data is not fully able to
resolve the bar flow.\footnote{Robust weighting of HI data yields
  velocity maps with higher spatial resolution but lower S/N.  For
  NGC~2903, however, these data from the bar region have too low S/N
  to allow meaningful fits.}  Fortunately, the velocity field of this
galaxy has also been mapped at higher spatial resolution in the
$^{12}$CO($J=1 \rightarrow 0$) line \citep{BIMA} and in H$\alpha$
using a Fabry-P\'erot instrument \citep{BHaB}.  We here show fits to
the H$\alpha$ data (kindly made available by Olivier Hernandez) since
they extend to larger radii than do the CO data, albeit with lower
spatial resolution ($4.8\arcsec$).

Fig.~\ref{n2903fig}(a) shows a 2MASS\footnote{Atlas Image obtained as
  part of the Two Micron All Sky Survey (2MASS), a joint project of
  the University of Massachusetts and the Infrared Processing and
  Analysis Center/California Institute of Technology, funded by the
  National Aeronautics and Space Administration and the National
  Science Foundation.} H-band image of the galaxy, together with (b)
the velocity map from the BH$\alpha$Bar survey, (c) our best fit model
and (d) \& (e) residuals from an axisymmetric and full bar flow fits.

Table \ref{n2903_pars} and Fig.~\ref{n2903}(b) give the results we
obtained from by applying \velfit\ to the H$\alpha$ data for NGC~2903
-- note that the radial scales in panels (a) \& (b) of
Fig.~\ref{n2903} differ.  We used a similar smoothing ($\lambda = 6
\times 10^{-5}$ in this case) in order to eliminate absurdly large
velocities in the bootstrap analysis, which introduces a bias, as
before, against bar angles close to zero; since this bias causes a
strong skewness in the distribution of bar angles, we estimated the
uncertainty in this quantity from only those values more negative than
the best fit value.  The parameters of the fit to these data are
generally in good agreement with those from the HI data, and the
projected bar angles differ within their uncertainties.

However, the H$\alpha$ data show a much more pronounced non-circular
flow pattern within the bar region ($R \la 60\arcsec$), with perturbed
velocities almost as large as \vmean, which in turn significantly
exceeds the estimated circular speed from an axisymmetric fit to the
same data (black line) and the derived \vmean\ from the HI data.  The
fitted velocities from the two datasets are in reasonable agreement
outside the bar region, for as far as the optical data extend.  The
uncertainties in the bar region are still large, and significantly
larger than the point to point variation in the best fit suggesting
that slightly more aggressive smoothing could be warranted.

Thus it is clear that inadequacies in the HI data are the reason
\TdBW\ concluded that non-circular motions within the bar were small.
While the beam size of the \things\ data is not fully adequate to
resolve the bar flow, the general paucity of neutral gas in the barred
regions of other galaxies suggests it is also likely that HI is simply
not a faithful tracer of the bar flow.  Data from a different
component of the ISM having better spatial resolution and fuller
spatial coverage do reveal a pronounced non-circular streaming
pattern, as expected for this strongly barred galaxy.

\subsection{Searching for Aspherical Halos}
We here attempt to constrain the shapes of dark matter halos by
searching for non-circular streaming motions in the outer discs of the
\things\ sample.  Since the current version of \velfit\ assumes the
galaxy plane to be flat, it cannot be used in warped regions, which
generally arise outside the visible disc.  We are therefore restricted
to just two galaxies in the sample, NGC~3198 \& NGC~2403, for which
the extended HI disc is known from the analysis of \cite{dBWB} to be
approximately coplanar with the inner disc.

Even though these two galaxies are not strongly warped, the analysis
of \TdBW, which allows changes in PA and $i$ from ring to ring, may
underestimate the ellipticity of the dark matter halos in the disc
plane.  However, the main advantage of \velfit\ over \reswri\ in these
circumstances is that it searches for a bisymmetric distortion that is
coherent over a wide range of radii and could, in principle, detect
very mild distortions that might be masked by various sources of
noise, such as turbulence and local spiral streaming.  Since it is a
more sensitive probe of halo shapes, it should either detect mild
distortions, if they are present, or place a tighter lower bound on
the axis ratio of the potential.

\subsubsection{NGC 3198}
\begin{figure}
\includegraphics[width=1.25\hsize,angle=270]{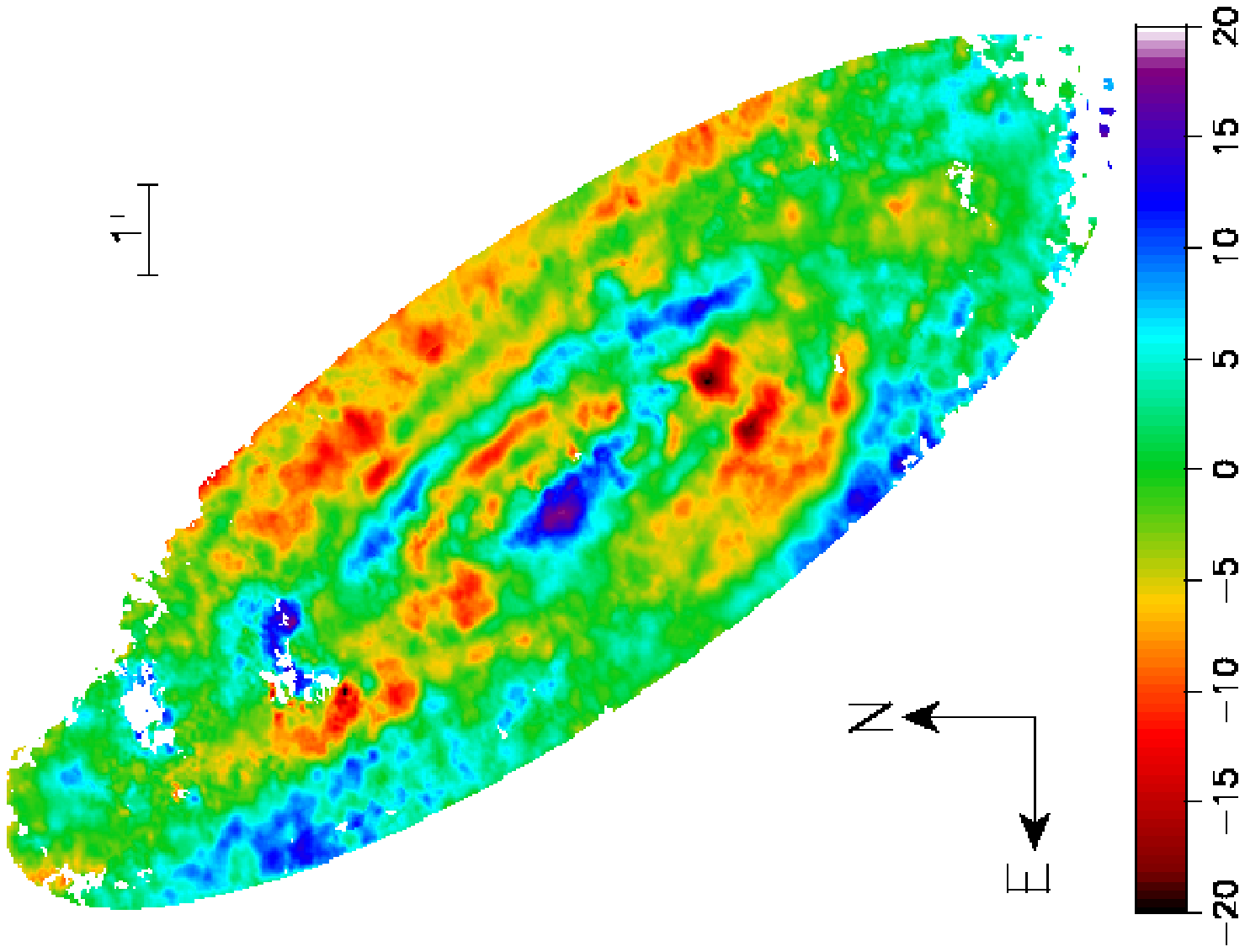}
\caption{Residuals after fitting an axisymmetric model to the
  \things\ data for NGC~3198.  The outer ellipse has a semi-major axis
  of $r=456\arcsec$ and velocities in the colour bar are in
  km~s$^{-1}$.}
\label{n3198_res}
\vspace{.5cm}
\centerline{\includegraphics[width=\hsize]{n3198_rc.ps}}
\caption{Results from \velfit\ using \things\ data for NGC~3198.  The
  black line shows the result of an axisymmetric fit to all radii,
  while the coloured points show $\vmean(R)$ (red), $V_{2,t}(R)$
  (green), \& $V_{2,r}(R)$ (blue) from a bisymmetric fit restricted to
  the range $240\arcsec \leq R \leq 460\arcsec$ with
  an axisymmetric model fitted to other radii.}
\label{n3198}
\end{figure}

Fig.~\ref{n3198_res} shows the residual map to $R=456\arcsec$ when a
flat, axisymmetric model is subtracted from the \things\ data for
NGC~3198.  The residual velocities are generally small, peaking at
$\pm 20\;$km s$^{-1}$, which is consistent with the small variations
in $i$ \& PA for $r>200\arcsec$ reported by \cite{dBWB}.  However, the
residual pattern reveals clear indications of mild spiral arm
streaming, even far outside the optical disc \citep[$R_{25} \simeq
  255\arcsec$ in the B-band,][]{dV91}.

\begin{table}
\caption{Best-fit parameters for NGC 3198 and NGC 2403}
\label{n3198_pars}
\begin{tabular}{@{}lrr}
                   & NGC 3198       & NGC 2403 \\
\hline
Adopted distance (Mpc)  &   13.8     &   3.2 \\
Systemic vel. (km s$^{-1}$)    &  $660.2 \pm 0.3$ &  $133.5 \pm 0.3$ \\
Disc inclination $i$    &  $ 70\fdg4 \pm 0\fdg3$ &  $ 64\degr \pm 1\degr$ \\
Disc PA, $\phi_d^\prime$ &  $216\fdg2 \pm 0\fdg5$ &  $123\fdg8 \pm 0\fdg6$ \\
Bar axis $\phi_b$       &  $ 46\degr \pm 14\degr$ &  $296\degr \pm 12\degr$ \\
Projected ``bar'' axis $\phi_b^\prime$ &  $236\degr \pm 5\degr$ &  $82\degr \pm 5\degr$ \\
\end{tabular}
\end{table}

In order to search for a possible mildly non-axisymmetric halo, we
tried fitting a bisymmetric model, with no spirality, to the outer
disc.  Such a model may be able to identify a weak bar flow that could
be buried in the spiral noise.  The parameters of our best fit model,
which includes a bisymmetric flow for $R \ga 220\arcsec$, are
listed in Table~\ref{n3198_pars}.  Our estimated values of $V_{\rm
  sys}$, $i$, \& PA are in excellent agreement with those given by
\cite{dBWB}.

Fig.~\ref{n3198} shows that the best fit non-axisymmetric velocities
are no larger than $\sim 6$\% of the circular speed and vary slowly
with radius.  If the perturbed velocities are caused by a
slowly-rotating, mild oval distortion of the halo in the outer parts
where the rotation curve is approximately flat, we should observe
$V_{2,t} \simeq 1.5V_{2,r}$ (eq.\ \ref{amp}).  In fact, the
coefficients are not in this predicted ratio at any radius and even
the signs differ over the inner part of the fitted range suggesting a
different origin for the perturbed velocities, such as spiral arm
streaming for which a fixed axis and slow rotation for the
perturbation are inappropriate assumptions.  Our fitted values of
$\phi_b$, $V_{2,t}(R)$, and $V_{2,r}(R)$ are merely those that achieve
the largest reduction in $\chi^2$ from the residual pattern shown in
Fig.~\ref{n3198_res}.  In particular, the PA of the fitted ``bar''
varies with radius as we sub-divide the fitted region, as one would
expect if the fit is picking up different fragments of the spirals.

Thus we are unable to identify a velocity pattern indicative of a
non-axisymmetric halo.  Whatever possible halo distortion may be
present, it is clearly still weaker than the mild spiral features we
do detect.  We therefore concur with \TdBW\ that the data from
NGC~3198 are consistent with this galaxy living in a perfectly round
halo.

Even though we do not have a firm detection of a bar-like distortion
in the halo, we can use the results in Fig.~\ref{n3198} to place a
lower bound on its ellipticity by making a rough estimate of the
maximum halo distortion that could be masked by the spiral
contribution.  We therefore suppose that the disturbed velocities are
due to a combination of a slightly elongated halo and a spiral
disturbance in the disc plane.  In this situation, the variation of
the total amplitude of the velocity distortions will be modulated by
the changing angle between the two as the phase of the spiral changes
with radius; it will be a maximum when the distortions are aligned and
a minimum when they are perpendicular.  (This is a conservative
assumption, since amplitude and phase variations could also be caused
by the changing distance from a resonance in just a single spiral,
with no second component of the disturbance potential.)

For NGC~3198, the combination $A_{\rm tot} = (V_{2,r}^2 +
V_{2,t}^2)^{1/2}$ varies from a minimum of $A_{\rm tot} =
1.25\;$km~s$^{-1}$ at 296\arcsec\ to $A_{\rm tot} =
11.10\;$km~s$^{-1}$ at 444\arcsec.  If this spans a full range of
possible phase differences (0 to $\pi/2$), and the spiral and halo
contributions are separately roughly constant, then $A_{\rm tot} = A_s
+ A_h$ at maximum, and $A_{\rm tot} = A_s - A_h$ at minimum.  Solving
for the separate spiral and halo amplitudes, we find $A_s \simeq
6.3\;$km~s$^{-1}$ and $A_h \simeq 5.0\;$km~s$^{-1}$.  Since $V_{2,t} =
1.5V_{2,r}$ for a mild, non-rotating distortion with a flat rotation
curve (eq.\ \ref{amp}), we expect $V_{2,r} \simeq A_h / 1.8$.  Using
the values $V_{2,r} \simeq 5 / 1.8\;$km~s$^{-1}$ and $\vmean =
140\;$km~s$^{-1}$ in formula (\ref{potshape}), we find $q_\Phi \ga
0.98$ and $q_\rho \ga 0.94$ as our estimated lower limits on the axis
ratios of the potential and density of the halo in NGC~3198.

\begin{figure}
\includegraphics[width=.8\hsize,angle=270]{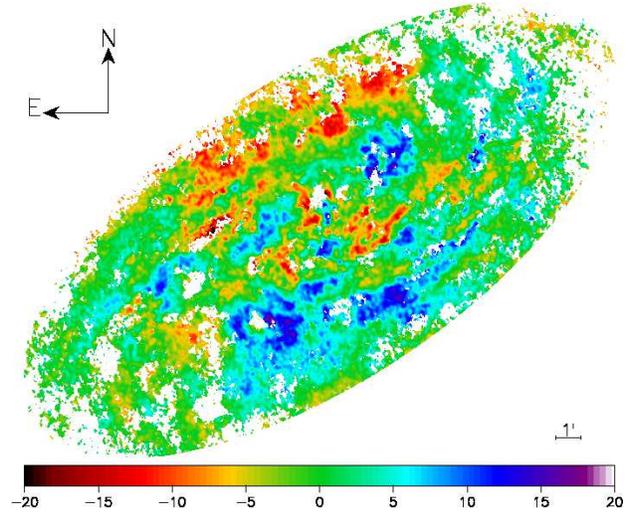}
\caption{Residuals after fitting an axisymmetric model to the
  \things\ data for NGC~2403.  The outer ellipse has a semi-major axis
  of $r=844\arcsec$ and velocities in the colour bar are in km~s$^{-1}$.}
\label{n2403_res}
\vspace{.5cm}
\centerline{\includegraphics[width=\hsize]{n2403_rc.ps}}
\caption{Same as for Fig.~\ref{n3198} but for NGC 2403.}
\label{n2403}
\end{figure}

\subsubsection{NGC 2403}
Fig.~\ref{n2403_res} shows that a simple, flat, axisymmetric model is
a good fit to the \things\ data for NGC~2403, consistent with the tiny
variations in $i$ \& PA reported by \citep{dBWB}.  The small residuals
are somewhat less indicative of spiral streaming than in NGC~3198, and
show hints of a more global N-S anti-symmetry at larger projected
radii \citep[$R_{25} \simeq 656\arcsec$ in the B-band][]{dV91}.

Our fitted values of $V_{\rm sys}$, $i$ \& PA from a model that
includes bisymmetric streaming velocities for $R>450\arcsec$, listed
in Table~\ref{n3198_pars}, are in excellent agreement with those found
by \cite{dBWB}.  Fig.~ \ref{n2403} shows that the perturbed velocities
are $\la 10\;$km/s, but the radial component is generally the larger,
which is inconsistent with a non-rotating, bar-like distortion.  Thus
we again concur with \TdBW\ that the \things\ data on this galaxy are
consistent with it living in a perfectly round halo.

While the perturbed velocities are not of the form expected for a
mild, non-rotating halo distortion, we proceed as for NGC~3198.  For
NGC~2403, we find $A_{\rm tot}$ varies from a maximum of $A_{\rm tot} =
11.86\;$km~s$^{-1}$ at 510\arcsec\ to $A_{\rm tot} = 2.12\;$km~s$^{-1}$
at 809\arcsec.  Again assuming this spans a full range of possible
phase differences and the spiral and halo contributions are separately
roughly constant, then $A_s \simeq 7.0\;$km~s$^{-1}$ and $A_h \simeq
4.9\;$km~s$^{-1}$.  Using the values $V_{2,r} \simeq 4.9 /
1.8\;$km~s$^{-1}$ and $\vmean = 120\;$km~s$^{-1}$ in formula
(\ref{potshape}), we again find $q_\Phi \ga 0.98$ and $q_\rho \ga
0.94$ as our estimated lower limits on the axis ratios of the
potential and density of the halo in NGC~2403.

\section{Conclusions}
\cite{SS07} devised the software tool \velfit, for fitting
non-circular streaming flows in galaxies caused by non-spiral like
distortions.  We have shown here that it is altogether superior to the
commonly used \reswri\ \citep{SFdZ}.  This is because it provides a
correction for the systematic error in the azimuthally averaged
orbital speed, it can fit strong distortions, it is more sensitive to
radially coherent disturbances, it readily allows radial smoothing of
the fitted velocities, and the estimated disturbed velocities are more
easily related to the potential distortion.  We also argue that while
the current version of \velfit\ assumes a flat plane for the inclined
disc, this is no more restrictive, since tilts of individual rings in
the bright inner part of the disk, as \reswri\ allows, are dangerous
and can lead to severe underestimation of the distorted velocities.
This paper illustrates these advantages for a number of galaxies.

We show that the \things\ data \citep{WBdB} for NGC~2976 can indeed be
fitted by an inner bar-like distortion, albeit somewhat milder than
that found by SS07 from H$\alpha$+CO data.  We argue that a bar flow
is more natural than either the radial flow fitted by \cite{SBLB} or
the twisted disk model of \citet{TdBW}.  As shown in SS07, the
improved mean orbital speed of the gas estimated from \velfit\ is a
fairer estimate than the simple mean fitted by tools such as \reswri.
The difference, which arises from the bias to the velocity on the
major axis, can be of either sign depending on the orientation of the
bar to the major axis of projection.

We also show that neutral hydrogen observations are not well suited to
tracing gas dynamics in strongly barred potentials.  The neutral
hydrogen generally has a low column density in the barred region, and
the velocity maps are generally too noisy or sparsely sampled to yield
a clear indication of bar flow.  Smoothing to lower spatial resolution
improves signal-to-noise, and yields an almost fully sampled velocity
map throughout the bar region of NGC~2903.  We identify an oval flow
pattern in these data of about the right physical size and with
signifiant streaming velocities, but the large uncertainties imply
they are also consistent with a round potential.  However, a strong
bar flow is unambiguously detected in our fit to the H$\alpha$
velocity map of \cite{BHaB}.  While the superior spatial resolution of
the optical data is clearly important, the generally patchy and faint
emission from neutral hydrogen in the barred regions of this and other
barred galaxies in the \things\ sample suggest that neutral hydrogen
is simply a poor tracer of bar flows.

Our analysis of the \things\ data for NGC~3198 \& NGC~2403 reaches a
similar conclusion to that of \TdBW: that the halos of these two
galaxies are close to round.  \cite{Jog00} and \cite{BSB07} show that
the self-consistent response of the disc can mask a large part of the
distortion in the halo, but only when the disc is massive.  In these
two cases, the outer gas disc probably has little mass and therefore
could not hide a more substantial halo distortion.

However, it is hard to confront this result with the predictions of
LCDM (cited in the introduction): not only is it based on just two
galaxies, but it is possible these two galaxies are unrepresentative
perhaps because only galaxies with unusually round halos could host an
extensive disc of gas that is not warped!  Clearly, measurements of
halo shapes in a representative galaxy sample will require a tool that
can reliably measure potential distortions in warped discs; we leave
development of such a capability for future work.  Other statistical
approaches \citep{FdZ92,TdBM} do, however, place some reasonably tight
constraints on halo shapes.

Thus, while we confirm the conclusion of \TdBW\ that many galaxies in
the \things\ sample have at most minor departures from axial symmetry,
\velfit\ reveals that mild bars are present in NGC~2976 and NGC~7793.
We also show that a pronounced non-axisymmetric flow is revealed more
clearly in other data for the strongly barred galaxy NGC~2903.

Another valuable application for \velfit\ will be a preliminary
analysis of the velocity maps of strongly barred galaxies.  It would
be very helpful to obtain a clear indication of whether the flow
pattern is, or is not, well enough sampled and sufficiently regular to
yield a result, before embarking on laborious mass modeling by the
method described by \cite{WSW01}.  Furthermore, such a study needs
approximate axisymmetric mass models, and the estimates of
\vmean\ from \velfit\ will be more useful than the ``circular speed''
estimated from other less powerful tools.

\section*{Acknowledgments}
We thank Claude Carignan and Olivier Hernandez for supplying
Fabry-P\'erot maps of several galaxies from their BH$\alpha$Bar
survey.  We also thank Andrew Baker, Ted Williams, and especially Tad
Pryor for discussions.  Kristine Spekkens, Jack Hughes, and an
anonymous referee provided comments that have helped us to improve
the paper.  This work was supported by NSF grant AST-0507323.

\appendix
\section{Applying a Smoothing Penalty}
The usual $\chi^2$ function is defined in eq.~(\ref{chisq}), but we can
minimize a new function if we wish
\begin{equation}
X^2 = \chi^2 + \lambda \sum_{k=2}^{K-1} {\cal A} \left[ V_{k-1} - 2V_k +
V_{k+1} \right]^2,
\end{equation}
which adds a penalty for large second differences between the
tabulated values $V_k$, which are assumed to be equally spaced.  The
constant ${\cal A}$, defined below, has dimensions of inverse square
velocity in order to fulfil the requirement that the smoothing penalty
be dimensionless.  The value of the smoothing parameter, $\lambda$,
which can be set independently for the axisymmetric and
non-axisymmetric terms, can be adjusted as desired; small values have
slight effect, while large values make the profile very smooth.
Over-smoothing not only increases the value of $\chi^2$, but may
reduce the inner slope and smooth other real features in the rotation
curve.

Since the magnitude of the second velocity difference (in square
brackets) varies as the inverse square of the ring spacing, $\Delta R
= R_{\rm max}/N_R$, we choose ${\cal A} = (R_{\rm max}/\Delta R)^4 /
V_{\rm typ}^2 = N_R^4 / V_{\rm typ}^2$, which ensures that the
smoothing constraint is unaffected when the ring spacing is
changed. Here, $V_{\rm typ}$ is a constant that is a rough estimate of
a typical orbital speed in the disc.

The smoothness constraint is applied only in the matrix that is solved
to find the optimal values for $V_k$ (see SS07), and adds extra terms
as follows:
\begin{eqnarray}
{\partial X^2 \over \partial V_j} \!\!\! & = & \!\!\! {\partial \chi^2
  \over \partial V_j} \\ \nonumber & & + 2\lambda\,{\cal A} \left(
V_{j-2} - 4V_{j -1} + 6V_j - 4V_{j+1}+ V_{j+2} \right).
\end{eqnarray}

\section{Bootstrap Estimation of Errors}
The bootstrap uses the spread of estimated quantities from repeated
fits to resampled (or pseudo-)data to yield a non-parametric estimate
of the true uncertainty in each quantity.  To construct one
realization of pseudo-data, we add to the predicted velocity from the
best-fit model at every pixel a residual velocity from some other
pixel chosen at random from residuals between the best fit model and
the real data.  Completely random resampling of residuals assumes
that the residuals are uncorrelated, whereas inspection of our
residual images generally reveals coherent patterns of residual
velocities due to features, such as other forms of non-axisymmetric
streaming and large-scale turbulence, which are not included in our
fitted model.  Pseudo-data constructed by fully random interchange
eliminates correlations between the residuals and generally leads to
unrealistically small uncertainties.  Thus we require a scheme that
will reproduce appropriately correlated errors at random at each
iteration of the bootstrap.

\cite{SS07} adopted constant residuals over small patches of the
image, which worked well for the small galaxy NGC~2976 where most of
the correlations appeared to arise from turbulence.  But that scheme
cannot capture the patterns of residuals that arise from spirals and
other non-axisymmetric distortions that are clear features in the
residuals for larger galaxies.  We therefore adopt a different
approach here.

Since the dominant residual correlations appear over azimuthally
extended regions in the disc plane, we manipulate the actual residual
pattern as follows: we deproject it to face on and then, at every
iteration, we rotate the residuals through a random angle.  We also
shift the residual pattern outwards by adding a constant to the radius
of every pixel and subtract the maximum radius from those pixels that
are shifted outside the map so that they fill the hole created in the
centre by the outward shift.  The constant used to shift in radius is
a randomly-chosen fraction of the radius of the map.  We then
reproject the new residual pattern, and assign a residual velocity at
each pixel in the pseudo-data from that of the nearest pixel in the
scrambled residual image.

When we fit for an axisymmetric model in part of the galaxy and
include a non-axisymmetric perturbation over a limited radial region,
we scramble the residuals within each of these parts separately, and
do not interchange residuals between the separate regions, since the
appearance of residual patterns in the two regions can differ quite
markedly.

\label{lastpage}

\end{document}